\documentclass{aa}  
\usepackage{graphicx}
\usepackage{txfonts}
\usepackage{natbib}
\bibpunct{(}{)}{;}{a}{}{,}

\begin{document}
   \title{Submillimeter to centimeter excess emission from the Magellanic Clouds.}
   \subtitle{II. On the nature of the excess}
   \titlerunning{mm-cm excess in the Magellanic clouds}

   \author{C. Bot \inst{1,} \inst{2}
          \and
          N. Ysard\inst{3}
          \and
          D. Paradis \inst{4}
          \and
          J.P. Bernard\inst{5}
          \and
          G. Lagache \inst{6}
          \and
          F.P. Israel \inst{7}
          \and
          W.F. Wall \inst{8}
          }

   \offprints{C. Bot}

   \institute{Universit\'e de Strasbourg, Observatoire Astronomique de Strasbourg
   	\and
	CNRS, Observatoire Astronomique de Strasbourg\\
	UMR7550, F-67000 Strasbourg, France\\
              \email{caroline.bot@astro.unistra.fr}
              \and
              Department of Physics, P.O. Box 64, FIN-00014 University of Helsinki, Finland
              \and
		Infrared Processing and Analysis Center, California Institute of Technology, Pasadena, CA91125, USA
		\and
		Universit\'e de Toulouse, UPS, CESR, F-31028 Toulouse, France
		\and
		Institut d'Astrophysique Spatiale, F-91405 Orsay, France
		\and
		Sterrewacht Leiden, Leiden University, P.O. Box 9513, 2300 RA Leiden, The Netherlands
		\and
		Instituto Nacional de Astrof\'isica, \'Optica y Electr\'onica, Apdo. Postal 51 y 216, Puebla, Pue., M\'exico
             }

   \date{Received ...; accepted ...}
 
  \abstract
   {Dust emission at sub-millimeter to centimeter wavelengths is often simply the Rayleigh-Jeans tail of dust particles at thermal equilibrium and is used as a cold mass tracer in various environments including nearby galaxies. However, well-sampled spectral energy distributions of the nearby, star-forming Magellanic Clouds have a pronounced (sub-)millimeter excess (Israel et al., 2010).}
   {This study attempts to confirm the existence of such a millimeter excess above expected dust, free-free and synchrotron emission and to explore different possibilities for its origin.}
   {We model near-infrared to radio spectral energy distributions of the Magellanic Clouds with  dust, free-free and synchrotron emission. A millimeter excess emission is confirmed above these components and its spectral shape and intensity are analysed in light of different scenarios: very cold dust, Cosmic Microwave Background (CMB) fluctuations, a change of the dust spectral index and spinning dust emission.}
   {We show that very cold dust or CMB fluctuations are very unlikely explanations for the observed excess in these two galaxies.  The excess in the Large Magellanic Cloud can be satisfactorily explained either by a change of the spectral index due to intrinsic properties of amorphous grains, or by spinning dust emission. In the Small Magellanic Cloud however, due to the importance of the excess, the dust grain model including TLS/DCD effects cannot reproduce the observed emission in a simple way. A possible solution was achieved with spinning dust emission, but many assumptions  on the physical state of the interstellar medium had to be made.
   }
   {Further studies, using higher resolution data from Planck and Herschel, are needed to probe the origin of this observed submm-cm excess more definitely. Our study shows that the different possible origins will be best distinguished where the excess is the highest, as is the case in the Small Magellanic Cloud.}


   \maketitle
%

\section{Introduction}

(Sub-)millimeter to centimeter emission in galaxies is thought to be quite simple, originating from a combination of dust emission, free-free and synchrotron emission. Thermal free-free radiation originates in the ionized gas of H{\sc ii} regions and its emission is nearly flat in this optically thin regime of the spectrum ($S_\nu\propto \nu^{-0.1}$). Synchrotron radiation is emitted by relativistic electrons accelerated in magnetic fields and its spectrum is also a power law ($S_\nu \propto \nu^{-\alpha_{sync}}$). The spectral index of synchrotron emission and the relative contribution of these two emission mechanisms to the intensity observed at a given wavelength is constrained by radio observations. Dust emission observed at far-infrared and millimeter wavelengths is produced by large dust grains in thermal equilibrium that emit as a modified black-body ($S_\nu\propto \nu^{\beta}B_\nu(T_{dust})$ where $\beta$ is frequently taken as $\sim 2$). Because of the wavelength dependence of thermal dust emission, millimeter dust emission is often used as a tracer of the cold molecular gas reservoir in a galaxy \citep[e.g.][]{GZM+93,Dunne:2001vn,RBR+04,Weis:2008kx,Vlahakis:2008yq}. This information is particularly interesting since cold H$_2$  is almost impossible to observe and the use of CO line emission is not straightforward, in particular at low metallicities \citep{Israel:1988uq,LLD+94,Israel:1997kx}. Millimeter dust emission could be a good alternative to the CO molecule for tracing cold matter, also in low metallicity environments \citep{Thronson:1988vn,Israel:1997ys,RBR+04,Bot:2007yq, Bot:2009hb}.

This simple picture is however thrown into doubt by observations of the Milky Way and nearby galaxies.  Several studies present excess emission in the sub-millimeter that at face value trace large amounts of very cold dust with temperatures below 10K \citep{Reach:1995jt, Chini:1995ao,Krugel:1998zm, Popescu:2002ix,GMJ+03, Dumke:2004wq}. But, as already noted by \citet{Reach:1995jt}, the spatial correlation of this very cold dust with the warm component in our galaxy argues against this interpretation. In nearby galaxies also, the scenario of very cold dust is rejected based on different physical arguments \citep[e.g.]{LIS+02}. This sub-millimeter excess is then alternatively attributed to variations of optical properties of dust grains. For example, low spectral emissivity index values, $\beta$, might apply \citep{DDB+03, ABC+03}. Variations of the spectral emissivity index with dust temperature have been noted \citep{DBB+03,Desert:2008ve}. These variations could be related to physical properties of amorphous solids \citep{Meny:2007wo}. However, in nearby galaxies, such variations in dust properties might be difficult to disentangle from the effect of mixing different dust components, with different dust temperatures, along the line of sight. Data at numerous wavelengths covering the far-infrared (FIR) to the millimeter regime are required to better constrain the emission from dust in nearby galaxies.

At longer wavelengths ($\sim 30$GHz, $\sim 1$cm), another excess of emission has been observed in our galaxy and is called ``anomalous dust'' \citep[e.g.][]{Draine:1998cr,Lagache:2003mk,de-Oliveira-Costa:2004xu, Finkbeiner:2004dz,Casassus:2006hi,Miville-Deschenes:2008bh,Dobler:2008ft}. This excess is best explained by small rotating dust grains \citep["spinning dust"][]{Draine:1998fh}. Anomalous dust emission remains poorly constrained by observations (in particular its variations with the environment). Theoretical models predict that the peak frequency of  spinning dust emission should shift with grain size or density \citep{Ali-Haimoud:2009tg,Ysard:2010hc}. What is the influence of spinning dust emission on submillimeter to radio spectral energy distribution of galaxies?

The Small and the Large Magellanic Clouds (\object{SMC} and \object{LMC}) are two of the nearest galaxies to ours (located at distances of ~60 and ~50kpc respectively \citep{Deb:2010ve,Szewczyk:2009qf,Schaefer:2008zr,Szewczyk:2008ly}). As such, they have been extensively observed in all wavelength regimes. Their low metallicities ($\sim 1/2$ and 1/6 for the LMC and the SMC respectively \citep{Pagel:2003bh}) provide an opportunity to test our understanding of astrophysical processes in different physical conditions from our galaxy. 

\citet{Israel:2010fk} build spectral energy distributions (SEDs) of the Magellanic Clouds by combining selected literature flux densities with COBE and WMAP data. The spatial resolution is low ($\sim 1^o$ beam), but the resulting SEDs have an unprecedented  sampling of wavelengths, particularly at the far-infrared to the radio regime and show a pronounced (sub-)millimeter excess. In this papier, we analyse these SEDs of the full SMC and LMC to assess the contribution of dust, free-free and synchrotron emission, in order to quantify and study this excess emission and its possible origin. Our study is a prototype of future work on more distant galaxies that will use Herschel (60-600$\mu$m) and Planck ( up to$\sim$10~mm) data.
 
In section \ref{sec:intsed}, we present a fit of the integrated spectral energy distributions of the Magellanic Clouds, at  infrared and radio wavelengths, with dust, free-free and synchrotron emission. Above these three components, a significant emission excess is observed in the millimeter-centimeter domain. We explore whether this excess could be associated with very cold dust (Sec. \ref{sec:vcd}), with cosmic microwave background fluctuations (Sec. \ref{sec:cmb}), with the amorphous nature of dust grains (Sec. \ref{sec:meny}), or with spinning dust emission (Sec \ref{sec:spinning}).

\section{Integrated spectra of the Magellanic Clouds \label{sec:intsed}}

\begin{figure*}
\includegraphics[width=0.5\textwidth]{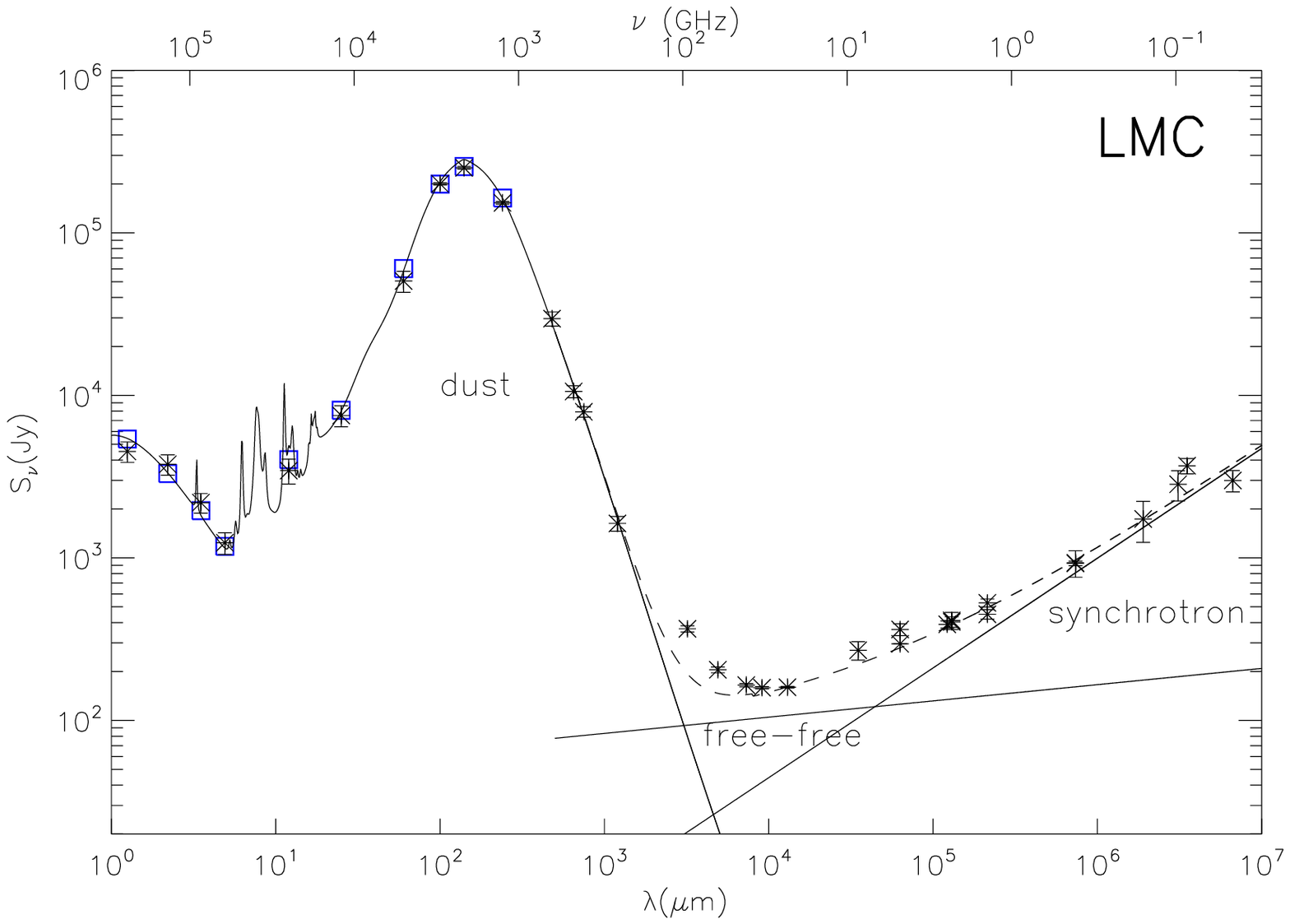}\hfill \includegraphics[width=0.5\textwidth]{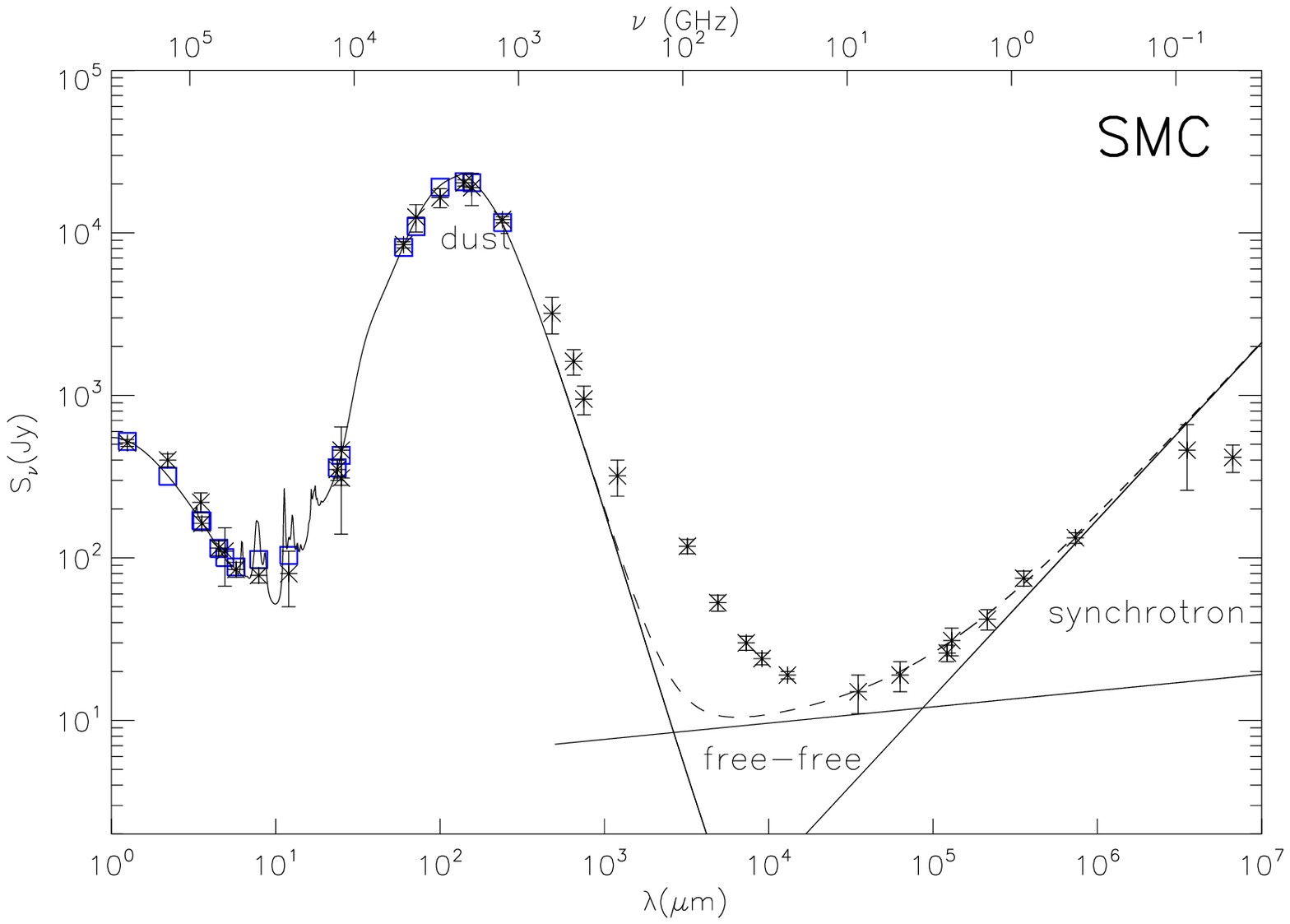}
\caption{ Spectral energy distributions from the IR to the radio for the integrated Magellanic clouds. The \citet{Draine:2007lr} dust model spectra as observed in the different photometric bands (blue squares) is fitted to the observed data points (black stars) from 1.27 to 240$\mu$m.  Synchrotron and  free-free spectra are adjusted on the WMAP and radio data. The sum of the three component fit is displayed as a dashed line. Error bars correspond to 1$\sigma$ uncertainties.\label{fig3}}
\end{figure*}

We used the spectral energy distribution determined by \citet{Israel:2010fk}  for the Small and the Large Magellanic Clouds. These integrated spectral energy distributions are shown in Figure \ref{fig3} and the infrared and radio emission is fitted with three components: the \citet{Draine:2007lr} dust model, a free-free component and synchrotron emission.

\pagebreak[2]

The choice of a model to represent dust in the Magellanic Clouds is still problematic \citep[e.g.][]{Meixner:2010dq,Gordon:2010kl,Roman-Duval:2010oq,Kim:2010cr, Hony:2010nx}. We chose to use the \citet{Draine:2007lr} dust model, which is well suited to studies of galaxies spectral energy distribution and has been applied to a number of nearby galaxies observed in the infrared and the submillimeter \citep{Draine:2007fk}. In this model, a mixture of dust grains is heated by a distribution of starlight intensities. A power law distribution of interstellar radiation field intensities, between $U_{min}$ and $U_{max}$, represents photo-dissociation regions (PDR), while a constant interstellar radiation field $U_{min}$ illuminates the diffuse medium of the galaxy. 

Different dust mixtures and dust size distributions exist in this model, among which three have been tailored to the LMC environment and one for the SMC \citep{WD01, LD02}. However, the SMC dust model was tailored to SMC extinction curves built on observations of only a few stars that might not represent the interstellar medium of the SMC as a whole \citep{Gordon:1998bs}. Indeed, fitting the "typical" SMC extinction curve (no 2175\AA bump, steep FUV rise), \citep{WD01} implied a lack of very small carbonaceous grains and a predominance of silicate dust grains in the SMC, which contrasts with the 8$\mu$m emission, the clear PAH features and the lack of silicate features (in absorption or in emission) as it is observed in many regions of the SMC \citep{Sandstrom:2010fk,Bot:2009hb}. Furthermore, studying 17 nearby galaxies observed from the near-infrared to the sub-millimeter, \citep{Draine:2007fk} showed that in no case the SMC bar dust model was preferred to the Milky-Way (MW) or LMC dust model.  Given these considerations, we chose to model the dust emission in the SMC with the \citet{Draine:2007lr} dust model using MW type dust. Note however that this choice should not significantly affect the far-infrared part of the spectrum. 

The LMC and SMC integrated fluxes from 1.27 to 240$\mu$m  were thus fitted with LMC dust models and  MW dust models respectively. To do the fit, we use the model spectra as observed with the DIRBE, IRAS and Spitzer photometric bands (i.e. taking into account the spectral response of the instruments and color corrections with respect to the reference spectral shapes) and compare them to the observed fluxes in these same bands, which avoids the use of color-corrections. The best model dust spectra obtained from the fitting procedure are presented in Figure \ref{fig3} and the corresponding parameters obtained are summarized in Table \ref{tab1}.

\begin{table}
\caption{Results from dust emission modeling of the integrated spectra: minimum and maximum radiation fields ($U_{min}$ and $U_{max}$, in units of the local interstellar radiation field), PDR fraction ($\gamma$), PAH mass fraction ($q_{PAH}$) and total dust mass (in $M_\odot$)}             
\label{tab1}      
\centering                          
\begin{tabular}{c c c c}        
\hline\hline                 
Parameter & LMC & SMC & SMC (with $U_{min}=0.1$) \\    
\hline                        
   model & LMC & MW & MW\\
   $U_{min}$&2.5& 2.0 & 0.1\\
   $U_{max}$& $10^3$& $10^3$ & $10^3$\\
   $\gamma$ &0.5\%& 7\% & 74\%\\
   $q_{PAH}$ & 2.4 & 0.47 & 0.47\\
   $M_{dust}$& $3.6 \cdot 10^6$& $2.9 \cdot 10^5$ &$1.1 \cdot 10^6$\\
\hline                                   
\end{tabular}
\end{table}

Free-free emission in the Magellanic Clouds is taken from the best estimates determined from the best fit of radio data by \citet{Israel:2010fk} (i.e. $S_{23GHz}^{free-free}=136$ and $12.5$Jy for the LMC and SMC respectively, with a $\nu^{-0.1}$ law, and synchrotron spectral indexes are $\alpha_{LMC}=-0.70$ and $\alpha_{SMC}=-1.09$.) 
The corresponding free-free and synchrotron emission for the LMC and SMC are shown in Fig. \ref{fig3}. 

The sum of the three fitted components (dust + free-free + synchrotron) is shown as a dashed line in Figure \ref{fig3}. For the LMC, the observed emission is well fitted from infrared to millimeter, as well as in the radio domain. TopHat fluxes at (sub-)millimeter  wavelength \citep[475$\mu$m to 1.22~mm][]{ABC+03} are well reproduced by the expected dust emission fitted to the infrared emission. We observe a small, but significant excess  (6-7$\sigma$)  emission in the first three WMAP bands at 3.3, 5 and 7.5~mm (90,60 and 40 GHz), above extrapolated models (dust, free-free and synchrotron). For the SMC integrated spectrum, a clear excess is observed in all TopHat and first four WMAP bands (i.e. from 476$\mu$m to 10~mm). This shows that in the SMC, the sub-mm part of the SED can not be well accounted for with dust modelling from the knowledge of the far-infrared peak emission (up to $240\mu$m) only.

\begin{figure}
\includegraphics[width=0.5\textwidth]{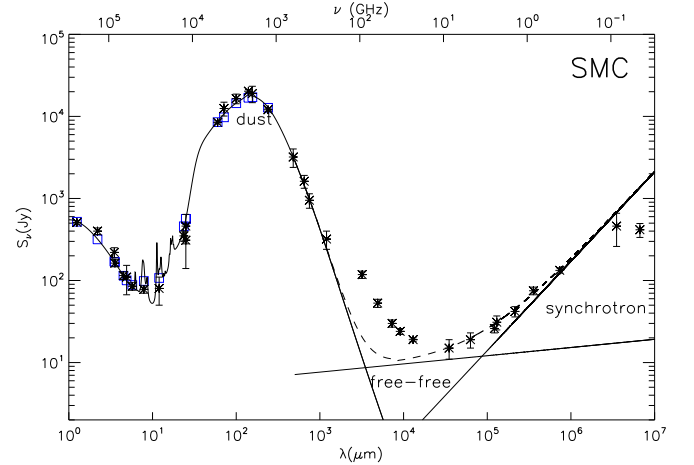}
\caption{Same as Fig. \ref{fig3}, right panel, but the minimum radiation field in the dust model is fixed to a value of 0.1 times the solar neighborhood value ($T_{dust}\sim 12K$). Error bars correspond to 1$\sigma$ uncertainties.\label{fig4}}
\end{figure}

Part of the excess observed in the SMC could be due to dust colder than can be extrapolated from  far-infrared data points. In particular, a $\sim 60\mu m$ excess has been reported in the Magellanic Clouds \citep{BBL+04,Bernard:2008zr}. This 60$\mu$m excess could be a significant part of the total emission in the SMC and bias dust models toward high temperatures. 
To check this, we performed a new fit of the SMC spectrum, similar to the modelling described above, but with a fixed minimum radiation field intensity: $U_{min}=0.1$ (which corresponds to $T_{dust}\sim 12K$ and is the lowest radiation field available in the pre-computed spectra given for \citet{Draine:2007lr} dust model). The result of the fit is shown in Figure \ref{fig4}. This model better reproduces observed dust emission, including TopHat data between 476~$\mu$m and 1.2~mm and will be chosen as our best model of dust emission in the SMC in order to be conservative. Still, a significant excess emission remains above the dust, free-free and synchrotron emission, between wavelengths of  3.3 and 10~mm.

\begin{figure}
\includegraphics[width=0.5\textwidth]{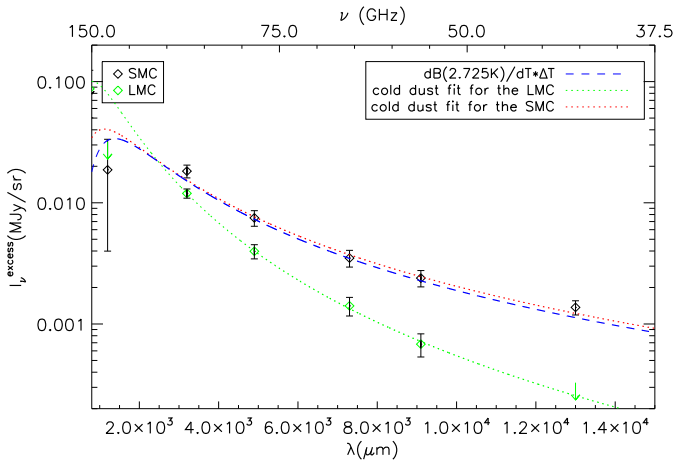}
\caption{Spectral energy distribution of the mm-cm excess in the LMC (in green) and in the SMC (in black). The spectral shape of the CMB fluctuations \citep[$dB_\nu(2.725K)/dT\times Delta T$][]{Fixsen:2009fk} is scaled to the SMC excess brightnesses for comparison. Modified black body fits to the LMC and the SMC excess  representing a possible cold dust component are shown as dotted lines. \label{fig5}}
\end{figure}

The spectrum of this excess in both galaxies is shown in Figure \ref{fig5} and Table \ref{tab2}. It is obtained by subtracting the best model emissions (dust, free-free and synchrotron) to the observed spectra. The spectral shape and surface brightness of the excess observed in the LMC and in the SMC are similar (c.f. Figure \ref{fig5}). This suggests that they probably arise from the same physical process. 

\begin{table*}
\caption{Surface brightness of the observed excess emission in the Magellanic Clouds. Error bars represent 1$\sigma$ uncertainties.\label{tab2}}             
\centering                          
\begin{tabular}{c c c c c c c c}        
\hline\hline                 
 &  I$_{1.2mm}$ & I$_{3.2mm}$ & I$_{5mm}$ & I$_{7.3mm}$ & I$_{9.1mm}$ & I$_{13mm}$ & $\Omega$ \\    
  & MJy/sr & MJy/sr & MJy/sr & MJy/sr & MJy/sr & MJy/sr & sr\\ 
\hline                        
SMC & $0.019\pm 0.014$ & $0.018\pm 0.002$ & $0.0075\pm 0.0011$ & $0.0035\pm 0.0006$ & $0.0024\pm 0.0004$ & $0.0014\pm 0.0002$ & 0.00544 \\
LMC & $<0.032$ & $0.012\pm 0.001$ & $0.0040\pm 0.0005$ & $0.0014\pm 0.0002$ & $0.0007\pm 0.0002$ & $<0.0003$ & $0.0160$ \\
 \hline                                   
\end{tabular}
\end{table*}

\section{CMB fluctuations?\label{sec:cmb}}

The intensity of the cosmic microwave background fluctuations are expected to be non-negligible at low surface brightnesses like the one observed for the Magellanic Clouds in the mm--cm range. 
Moreover, in the SMC,  the spectral energy distribution of the excess is of similar shape to the CMB fluctuations (c.f. Figure \ref{fig5}). 
We therefore estimate the probability  that CMB fluctuations in the background of the Magellanic Clouds are at the origin of this submm--cm excess.

\begin{figure*}
\includegraphics[width=0.5\textwidth]{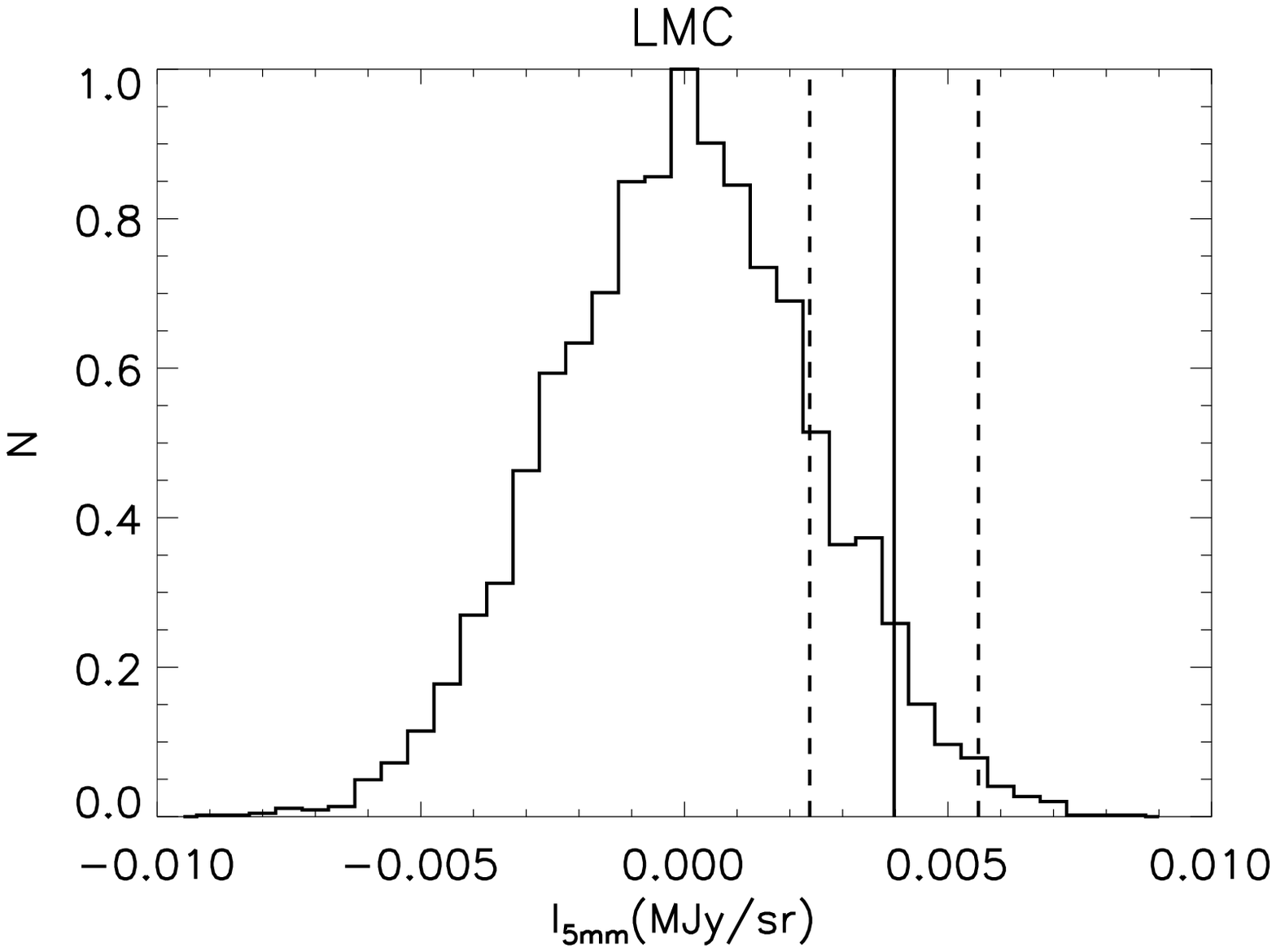}\hfill \includegraphics[width=0.5\textwidth]{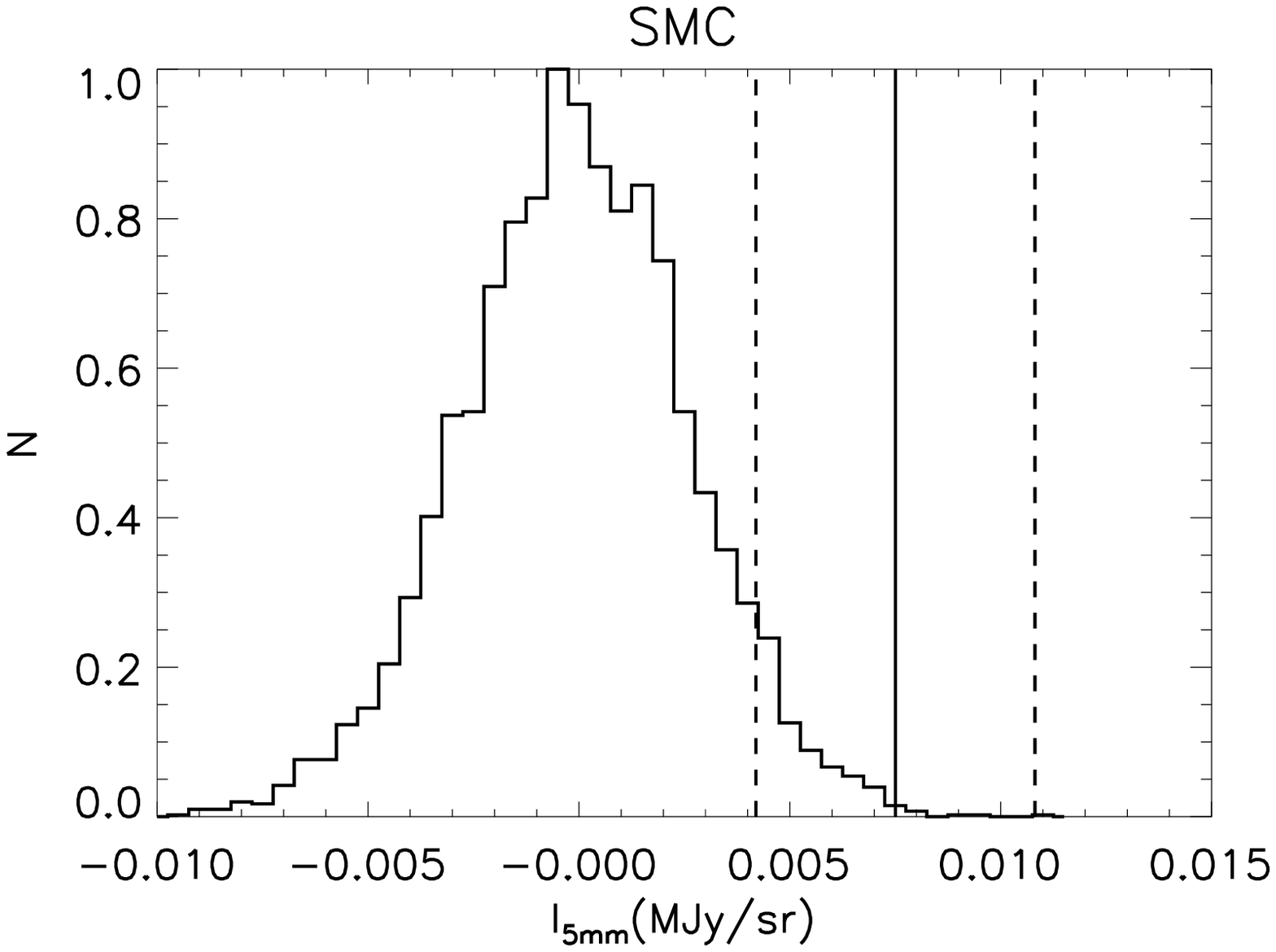}
\caption{Histogram of the simulated intensity fluctuations of the CMB measured similarly to the fluxes of the LMC and SMC (left and right panels, respectively) at 5mm (60 GHz), for 5000 random positions on the sky. The histograms have been normalized to their peak value and the bin size is $5\times 10^{-4}$MJy/sr). The solid vertical line in each pannel corresponds to the measured excess intensity at this wavelength, while the dashed lines give the 3$\sigma$ error estimates to either side of this measured excess. \label{fig7}}
\end{figure*}

To do so, we created a CMB simulation map that reproduces the observed CMB power spectrum \citep{Larson:2010fk}. We draw 5000 random positions in this CMB map and extract intensities of CMB fluctuations at 5mm (60GHz) inside an area equal to the one of the SMC and with an annular background subtraction. In Figure \ref{fig7}, we plot an histogram of CMB fluctuations intensities obtained for these 5000 random regions.  We then estimate the probability of observing CMB fluctuations  at a surface brightness that is compatible with the observed excess surface brightness (c.f. Tab. \ref{tab2}) at 5mm  within $ 3\sigma$ . We find that this probability amounts to 5\% for SMC.

We did the same estimate for the LMC. The histogram of CMB fluctuations intensities obtained with the same method is shown in Figure \ref{fig4} (left panel). The difference observed between the histograms for the LMC and the SMC corresponds to a size difference between the regions considered in each case ($\sim4^o$ and $\sim 2^o$ radii respectively). For the LMC, we find a 14\% probability that the intensity of the excess at 5mm is due to CMB fluctuations. However, the spectral shape of the excess in this case differs from the one of CMB fluctuations.

Furthermore, the fact that both Magellanic Clouds show a similar mm--cm excess emission despite their different size on the sky argues against the hypothesis that this excess is due to CMB fluctuations. 

We therefore consider it very unlikely that the excess in the Magellanic Clouds originates from CMB fluctuations in the background of these galaxies. We note that our test is based on average surface brightnesses and as such does not make any use of the spatial distribution of the excess in the Magellanic Clouds. A study of the excess with respect to the expected CMB fluctuations at different spatial scales would help to better discriminate the significance of CMB fluctuations in that respect.

\section{Very cold dust?\label{sec:vcd}}

It is tantalizing to try to explain the excess observed with very cold dust, especially since colder dust in the SMC can explain the sub-millimeter part of the excess (c.f. Section \ref{sec:intsed}).
To check this hypothesis, the excess observed is fitted by a modified blackbody representing a possible cold dust component. Best fits are shown in Fig. \ref{fig5} and give dust temperatures of $\sim 3$K and spectral indices of $\beta_{LMC}^{cold}=1.2$ and $\beta_{SMC}^{cold}=0.33$. Such a cold thermal equilibrium temperature for big grains is unrealistic. Furthermore, the spectral dust emissivity index deduced for the SMC is also unrealistically low. Indeed, current observations suggest $\beta$ to be in the range 1--2.5\citep{BAB+96,LAB+99,DBB+03,Paradis:2009kx} and the Kramers-K\"onig equations suggest that a value of 1 as a lower limit to $\beta$ \citep{Emerson:1988ij}.  The existence of very cold dust that would explain the observed millimeter excess therefore seems very unlikely.

\section{A change of the dust emissivity spectral index? \label{sec:meny}}

Another hypothesis for explaining the millimeter excess is a change of the spectral emissivity index from the FIR to the mm. Such a change has been observed in our Galaxy \citep{Paradis:2009kx} and is related to intrinsic dust properties. To test this assumption, we used a model proposed by \citet{Meny:2007wo}, based on the properties of amorphous dust grains. They consider two complementary physical models. The first one describes excitation of acoustic lattice vibrations, due to the interaction between electromagnetic fields and a disordered charge distribution (DCD).  This DCD model is characterized by a correlation length, used to describe the disorder degree of amorphous state.The second model focuses on the processes associated with a distribution of low-energy two-level systems (TLS), such as resonant absorption and relaxation processes, temperature-dependent emission. This DCD/TLS model was succesfully used to reproduce the FIRAS/WMAP spectrum of galactic diffuse emission \citep[][]{Paradis:2007eu,Paradis:2010fk}.

\begin{figure*}
\includegraphics[width=0.5\textwidth]{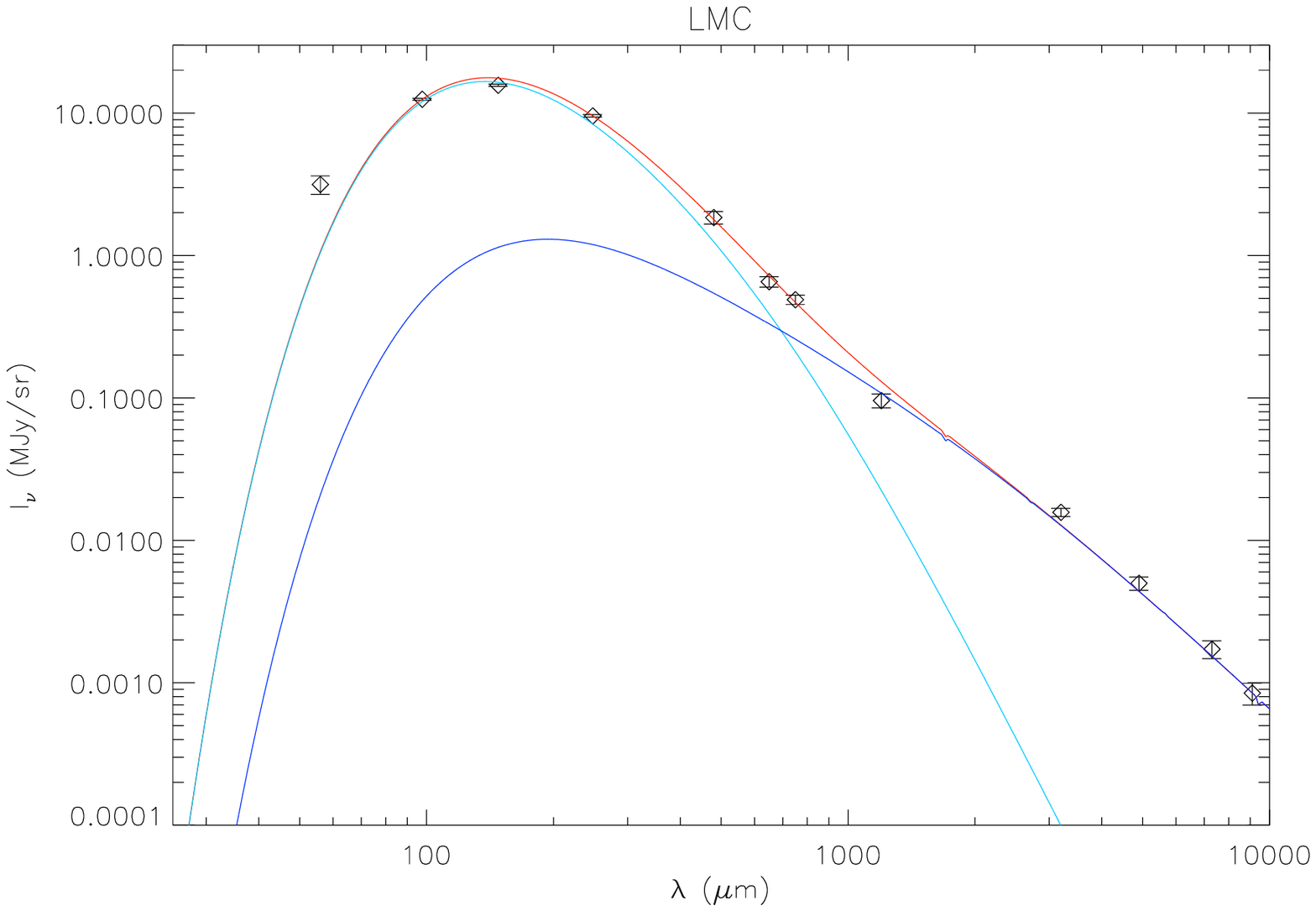}\hfill \includegraphics[width=0.5\textwidth]{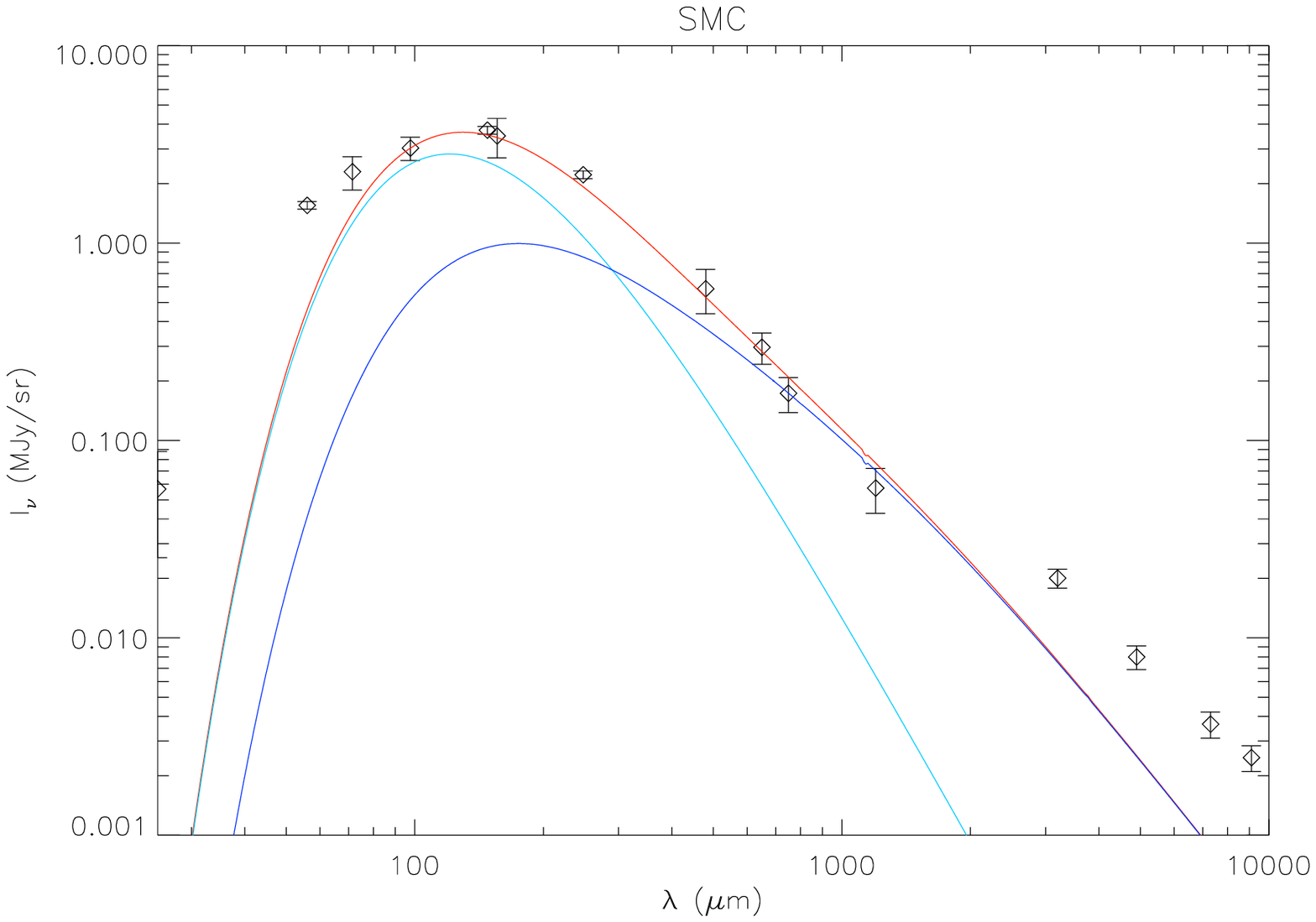}
\caption{SEDs of the Magellanic Clouds in the FIR to mm regime, once the free-free and synchrotron emission removed. This spectral energy distribution is fitted from 100$\mu$m to 9.1mm with a dust model with DCD/TLS effects.  The spectral shape of the DCD effect is shown in light blue while the TLS effects are displayed in dark blue and the total emission is displayed in red}.\label{fig8}
\end{figure*}

Fits to the LMC and SMC integrated spectra were done for all points from 100$\mu$m to $9.1$mm. To do so, three parameters were set to be free: the dust temperature, the correlation length ($L_C$) and the ratio of TLS over DCD effects ($A$ ratio). We did not attempt to fit the 60$\mu$m data point as the emission from very small grains that are  transiently heated is known to be significant at that wavelength. Best fits are shown in Figure \ref{fig8} and correspond to dust equilibrium temperatures of 21K and 24K, correlation lengths of 6.4 and 12.85 nm and $A$ ratios of 4.4 and 19.4, for the LMC and the SMC respectively. 

In the LMC, the DCD/TLS model is able to reproduce observations, assuming a decrease of the correlation length by a factor of 2 and an increase of the TLS processes intensity by a factor of 1.8, with respect to the parameters used to reproduce the spectrum of galactic diffuse emission. This result indicates a good agreement between dust properties in the diffuse medium of our Galaxy and in the overall LMC, even if the amorphization degree\footnote{The amorphization degree indicates if the internal structure of a grain is fully or partially amorphous.} of dust grains is different.

However, for the SMC, the flattening of the spectrum is too pronounced to be explained only by intrinsic dust properties.  Figure \ref{fig8}  shows a fit of the SMC SED with the DCD/TLS model. In this case, TLS effect intensity had to be enhanced by a factor of 8. Despite this enhancement, an additional component in the millimeter domain is needed to reproduce the data.

Hence, taking into account the properties of amorphous grains satisfactorily explains the excess in the LMC but the model fails at reproducing the one observed in the SMC.

\section{Spinning dust emission? \label{sec:spinning}}

The main mechanism invoked to explain Galactic anomalous excess at millimeter wavelengths is spinning dust emission, as first proposed by \citet{Draine:1998fh}. Unlike the excess in the Magellanic clouds, Galactic excess peaks between 20 and 40 GHz. However, models predict that the peak frequency of spinning dust emission shifts with radiation field intensity, interstellar gas physical parameters\footnote{Collision with interstellar gas species may excite or damp the rotation \citep{Draine:1998fh}.},  grain size distribution and electric dipole moment \citep{Ali-Haimoud:2009tg,Ysard:2010hc}. 

Here, we explore the possibility that the excess observed in the Magellanic Clouds originates from spinning dust emission produced by very small dust grains (i.e. PAHs). To do so, we fit the excess observed in the Magellanic Clouds with a model of spinning dust.  In both cases, and in order to reduce the number of free parameters, we make some ad-hoc assumptions. First, the electric dipole moment distribution is $\mu = m \sqrt{N}$ \citep{Draine:1998fh}, where $N$ is the number of atoms in the grain and $m$ is a constant equal to 0.4 D \citep{Ysard:2010hc}. The size distribution of dust grains is taken to be the same as in the \citet{Draine:2007lr} model (in order to be consistent with the fit of the mid-infrared part of the SEDs). We make the assumption that the interstellar medium of the Magellanic Clouds can be modelled as the sum of a diffuse medium and a PDR phase. The diffuse medium is assumed to have the same gas density as in the cold neutral medium in our Galaxy ($n_H \sim 30$ cm$^{-3}$), while the PDR component has the same density as the Orion Bar ($n_H \sim 10^4$ cm$^{-3}$). Using the radiation field distribution described in Tab. \ref{tab1}, other relevant gas parameters are calculated at thermal equilibrium with CLOUDY \citep{Ferland:1998fv}.  We assume $Z_{SMC} = Z_{\odot}/6$ and $Z_{LMC} = Z_{\odot}/2$. Parameters were taken from the optically thin zone of isochoric simulations with CLOUDY. The main gas species abundances vary with radiation field and density. We present some of the gas parameters obtained in Table \ref{tab3} for the different phases in each galaxy and for a sample of radiation field intensities.

\begin{table}
\caption{Gas parameters obtained with CLOUDY. Densities are expressed in $cm^{-3}$.}             
\label{tab3}      
\centering                          
\begin{tabular}{c | c c c c c}        
\hline\hline                 
 medium & G$_0$ & n$_H$ & $T_{gas}$ &n$_{C^+}$ & n$_{H^+}$\\    
\hline                        
\hline
LMC diffuse & 2.5 & 30 & 189.7 & $2.0\times 10^{-3}$ &  $6.8\times 10^{-3}$\\
\hline
LMC PDR & 2.5 & $10^4$ & 109.9 & $0.24$ &  $0.75$\\
 & 1000 & $10^4$ & 288.8 & $0.66$ &  $1.8$\\
 \hline
 \hline
 SMC diffuse & 0.1 & 30 & 124.1 & $6.5\times 10^{-4}$ & $3.6\times 10^{-3}$\\
 \hline                                   
 SMC  & 0.1 & $10^3$ & 115.2 & $3.7\times 10^{-3}$ & $3.8\times 10^{-2}$\\
moderate PDR & 1 & $10^3$ & 127.7 & $1.9\times 10^{-2}$ & $8.0\times 10^{-2}$\\
 & 1000 & $10^3$ & 1168.2 & $2.2\times 10^{-2}$ & $0.88$\\
 \hline                                   
 SMC  & 0.1 & $10^5$ & 63.2 & $3.6\times 10^{-4}$ & $0.3$\\
dense PDR & 1 & $10^5$ & 124.0 & $2.1\times 10^{-2}$ & $1.4$\\
 & 1000 & $10^5$ & 278.7 & $2.2$ & $9.3$\\
 \hline                                   
\end{tabular}
\end{table}

\begin{figure*}
\includegraphics[width=0.5\textwidth]{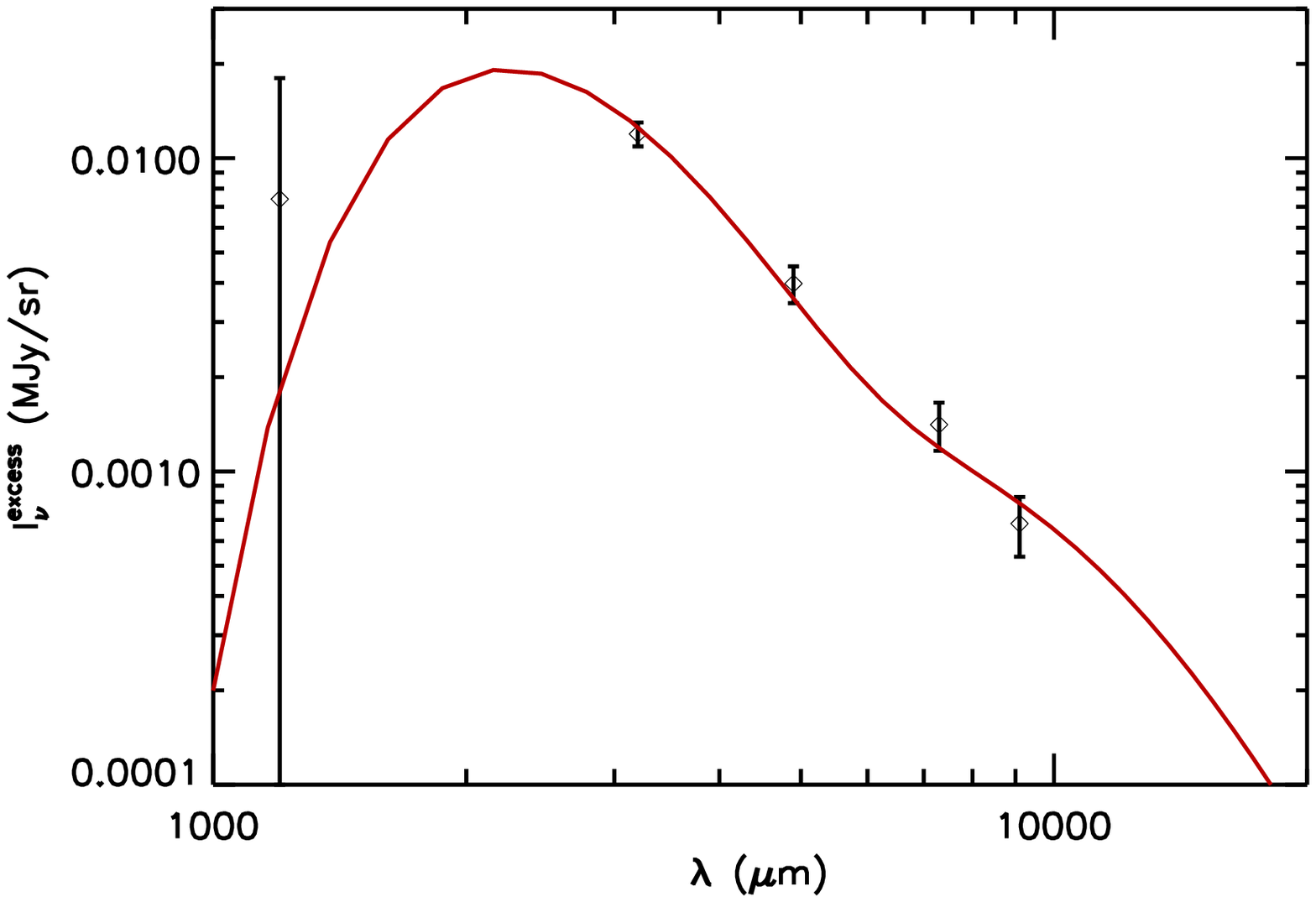}\hfill \includegraphics[width=0.5\textwidth]{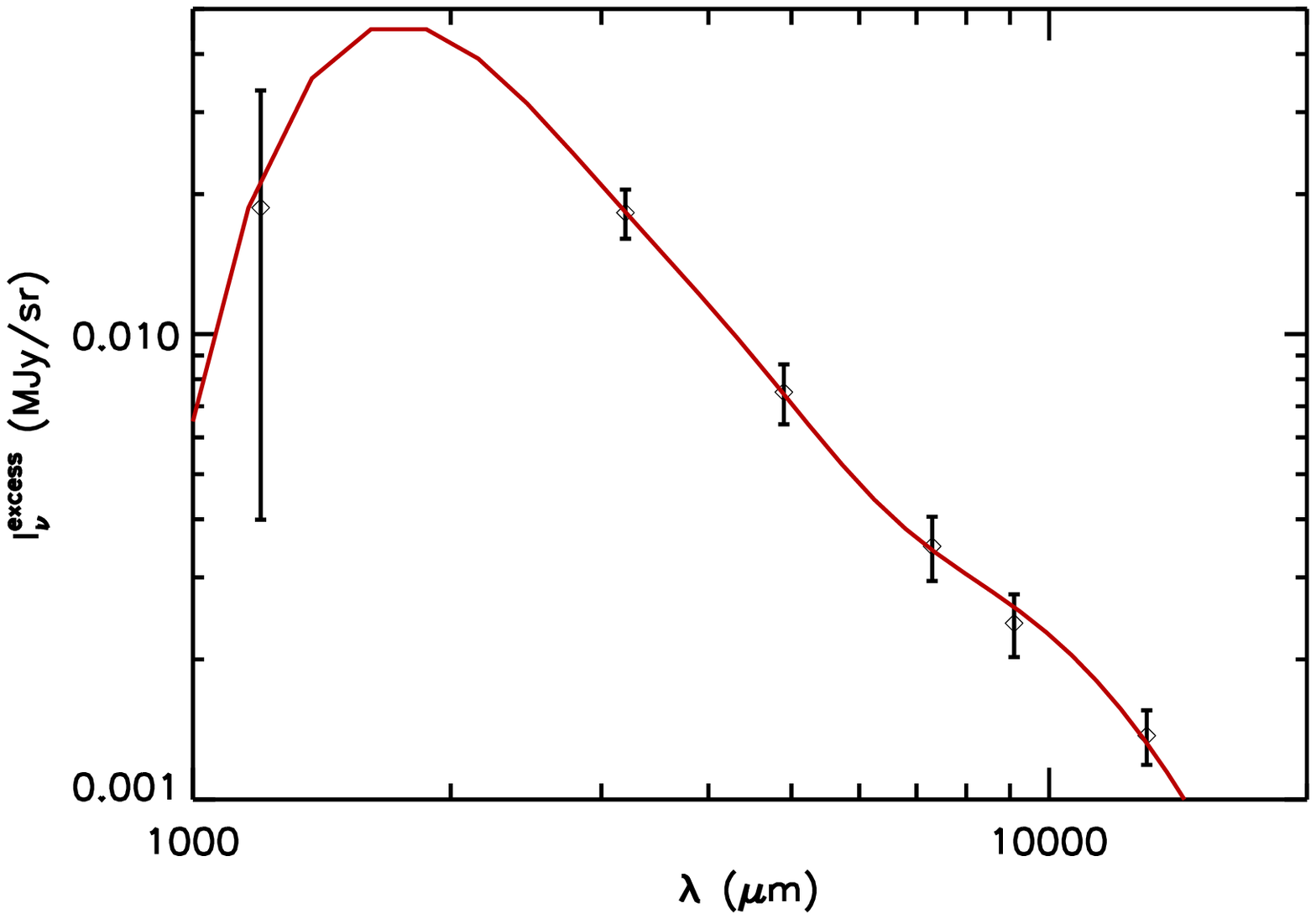}
\caption{Best fits of the millimeter excess observed in the Magellanic Clouds with anomalous dust components as described in the text. \label{fig9}}
\end{figure*}

For the LMC, a good fit is achieved and is shown in Fig. \ref{fig9} (left panel) overlaid on the observed excess emission. The resulting 
PAH mass fraction is $q_{PAH} = $2.15\%, with a carbon abundance in PAH of 0.9 ppm in the diffuse medium and 3.06 ppm in the PDR.

For the SMC however, no good fit was achieved with a diffuse+PDR model. Separating the PDR component into a dense PDR component (with $n_H=10^5$cm$^{-3}$) and a moderately dense PDR component (with $n_H=10^3$cm$^{-3}$), a good fit of the excess emission observed in the SMC is achieved and is shown in Fig. \ref{fig9} (right panel). In this case, we deduce $q_{PAH} = $ 0.39\%, with a carbon abundance in PAH of 2.20 ppm in the diffuse medium and 0.49 ppm in the dense PDR and 0.47 ppm in the moderately dense PDR. 

The spinning dust emission model can therefore reproduce the excess observed in both Magellanic clouds with parameters that are consistent with our current understanding of the interstellar medium in these environments. In particular, the PAH mass fractions obtained from the best fits of the millimeter excess are consistent with those obtained from the far-infrared dust emission modelling of the spectral energy distributions. This is consistent with the idea that spinning dust emission is produced by the smallest grains which are responsible for the mid-IR emission. 

If the millimeter-centimeter excess is due to spinning dust, the peak frequency of spinning dust emission in the Magellanic Clouds (139 and 160 GHz -- 2.2mm and 1.9mm-- for the LMC and SMC respectively) is shifted with respect to what is observed in our Galaxy. This shift depends on three main parameters: the size (the smaller the grain, the higher its emission frequency), the density (if the medium is denser, more grain-gas collisions occur and the anomalous emission peak is shifted toward higher frequencies) and the intensity of the Interstellar radiation Field (ISRF) (above a threshold of $G_0=10$, if the ISRF intensity increases, the emission peak is shifted to higher frequencies). In order to reproduce the millimeter excess observed in the Magellanic Clouds, all three parameters were equally important.

This study shows that spinning dust emission is a possible solution for the nature of the observed excess, in the two Magellanic Clouds. However, without complementary informations, many assumptions had to be made (e.g. mixing of two or three components along the line of sight, density of the interstellar medium). Also, recent studies \citep{Hoang:2010fk, Silsbee:2010uq} have shown that refinements on spinning dust models can lead to significant changes in the shape and in intensity of the spinning dust emission peak. The inclusion of such effects are beyond the scope of this paper, but could help the understanding of the excess in the SMC in terms of spinning dust emission.
We do not claim that the best fits obtained are a unique solution. Higher resolution studies with Planck and Herschel will be necessary to  confirm this origin of the excess by assessing the distribution of the millimeter excess in the Magellanic Clouds as a function of local conditions in better defined environments (e.g. diffuse medium only, HII regions). For example, it would be interesting to try to correlate the mm/cm excess in the Magellanic Clouds with PAHs emission to test the scenario of spinning dust emission.

\section{Conclusion\label{sec:conclusion}}

Comprehensive spectral energy distributions of the Magellanic Clouds from the near-infrared to the radio are modelled with dust, free-free and synchrotron emission. An excess above these three components is observed at millimeter wavelengths. This excess has a similar spectral shape in both Magellanic Clouds but is more prominent in the SMC total SED than in the LMC. 

We explore different scenarios for the origin of this excess emission in the millimeter regime: very cold dust, CMB fluctuations, a change of the dust emissivity spectral index and spinning dust emission. It is shown that very cold dust grains imply a thouroughly unrealistic nature for this excess and CMB fluctuations are unlikely to solely create such an excess.

For the LMC, the mm-cm excess can be explained equally well with TLS/DCD effects in amorphous grains or with spinning dust emission.

This is drastically different in the SMC, where the mm-cm excess is more pronounced. In this case, CMB fluctuations become very unlikely and the inclusion of TLS/DCD effects alone does not reproduce the excess. However, spinning dust emission models are capable to explain the observed excess with parameters that are consistent with our understanding of the ISM in the SMC, but many assumptions had to be made to obtain this result.

We hence confirm the existence of an unexpected mm-cm excess in the Magellanic Clouds, but the nature of this excess remains unclear. Spinning dust is a promising solution, but this should be further tested in better defined physical environments. The mm-cm excess observed in the Magellanic Clouds could easily be of a more complex nature, having multiple causes rather than a single one, so that the different effects explored here might add up. In this context, Planck and Herschel observations of the Magellanic Clouds  are needed to map this excess at a better spatial resolution, trace its spatial variations and probe its origin more definitely.

\begin{acknowledgements}
We would like to thank Nicolas Ponthieu for his help on the simulations of the CMB fluctuations.
\end{acknowledgements}

\bibliographystyle{aa}
\bibliography{../../../biblio}

\end{document}